\documentclass[wp,jeeabib,mathptmx,11pt,a4paper]{jeea}
\usepackage[utf8]{inputenc}
\usepackage{mathptmx}	
\usepackage{varioref}
\usepackage[english]{babel}
\usepackage{color,hyperref}
    \definecolor{linkcolour}{rgb}{0,0.1,0.4}
    \definecolor{citecolour}{rgb}{0,0.1,0.4}
    \definecolor{urlcolour} {rgb}{0,0.1,0.4}
    \hypersetup{colorlinks=true,linkcolor=linkcolour,citecolor=citecolour,filecolor=urlcolour,urlcolor=urlcolour}
\usepackage{graphicx}
\usepackage{epstopdf}
\usepackage{gensymb}
\usepackage{pdflscape}
\usepackage{array}
\usepackage{comment}
  \usepackage{booktabs}

  \newtheorem{prop}{Proposition}
  \newtheorem{corollary}{Corollary}

\newcolumntype{H}{>{\setbox0=\hbox\bgroup}c<{\egroup}@{}}

\makeatletter
\def\@fnsymbol#1{\ensuremath{\ifcase#1\or *\or \dagger\or \ddagger\or
    \mathsection\or 
    { \:\circ} \or
    **\or 
    \dagger\dagger
    \or \ddagger\ddagger \else\@ctrerr\fi}}
\makeatother

\Title{Optimal Pricing for Carbon Dioxide Removal Under Inter-Regional Leakage}
\Runninghead{Franks et al.}{Optimal Pricing for CDR untder Inter-Regional Leakage}

\ifdefined\DOUBLEBLIND
  \Author{---}
\else
\Author{Max}{Franks}{%
    Potsdam Institute for Climate Impact Research, Germany; 
    Technische Universität Berlin, Germany}{franks@pik-potsdam.de}
\Author{Matthias}{Kalkuhl}{%
    Mercator Research Institute on Global Commons and Climate Change, Berlin, Germany;
    Faculty of Economics and Social Sciences, University of Potsdam, Germany}{kalkuhl@mcc-berlin.net}
\Author{Kai}{Lessmann}{%
    Potsdam Institute for Climate Impact Research, Germany;
    Mercator Research Institute on Global Commons and Climate Change, Berlin, Germany}{lessmann@pik-potsdam.de}
\fi

\ifdefined\DOUBLEBLIND
\else
\Thanks{The authors gratefully acknowledge funding by the Deutsche Forschungsgemeinschaft (DFG, German Research Foundation) for project number 456947458.
The authors benefited from discussion with our colleagues Ottmar Edenhofer, Beatriz Gaitan, Hendrik Schuldt and Emilie Rosenlund Soysal at numerous workshops. We are grateful for their input.}
\fi

\Abstract{
\noindent Carbon dioxide removal (CDR) moves atmospheric carbon to geological or land-based sinks. In a first-best setting, the optimal use of CDR is achieved by a removal subsidy that equals the optimal carbon tax and marginal damages.  We derive second-best  policy rules for CDR subsidies and carbon taxes when no global carbon price exists but a national government implements a unilateral climate policy.
We find that the optimal carbon tax differs from an optimal CDR subsidy because of carbon leakage and a balance of resource trade effect. First, the optimal removal subsidy tends to be larger than the carbon tax because of lower supply-side leakage on fossil resource markets. Second, net carbon exporters  exacerbate this wedge to increase producer surplus of their carbon resource producers, implying even larger removal subsidies. Third, net carbon importers may set their removal subsidy even below their carbon tax when marginal environmental damages are small, to appropriate producer surplus from carbon exporters.
}


\begin{document}

\maketitle

\noindent This article is published as  
Max Franks, Matthias Kalkuhl, Kai Lessmann,
Optimal pricing for carbon dioxide removal under inter-regional leakage,
\emph{Journal of Environmental Economics and Management},
2022,
102769,
ISSN 0095-0696,
\hyperlink{https://doi.org/10.1016/j.jeem.2022.102769}{https://doi.org/10.1016/j.jeem.2022.102769}.

\section{Introduction}

Carbon dioxide removal (CDR) refers to a set of technologies that remove CO$_2$ from the atmosphere and store it in geological, terrestrial, or ocean reservoirs, or in products. While CDR is not employed at large scale today, the technology is projected to play a substantial role in achieving the Paris climate targets \citep{ipcc_global_2018}. An extensive literature has investigated its potentials, costs and side effects \citep{minx_negative_2018,fuss_negative_2018,nemet_negative_2018}, but research into efficient governance of CDR is in its infancy \citep[but cf.][]{Lemoine2020, GroomVenmans2021, KalkuhlFranksGrunerEtAl2022}. 

If removed carbon can be stored permanently, then reducing CO$_2$ emissions and employing CDR are two mitigation options that have the same effect on CO$_2$ in the atmosphere. Both can be incentivized by carbon pricing. Under idealized conditions, cost-benefit analysis mandates both prices to be equal to the social cost of carbon.
However, the general optimality condition may not hold in second-best settings such as unilateral carbon pricing in an international context as explored in this paper.  

In this paper, we reveal an asymmetry between carbon taxes and CDR subsidies for the case of supply-side carbon leakage and strategic appropriation of the surplus of resource producers.
Carbon leakage
refers to global emission responses to unilateral emissions reductions.  There are three commonly distinguished channels
\citep[following][]{JakobSteckelEdenhofer2014}:    
free-rider leakage, i.e.\ increased emissions in response to reduced climate change damages (as in \citealt{Hoel1991}), supply-side leakage, i.e.\ increased fossil energy demand in response to falling international energy prices (as in \citealt{Bohm1993, GerlaghKuik2014}), and specialization leakage, i.e.\ relocation of emission-intensive production (as in \citealt{siebert_environmental_1979}). Our analysis focuses on the first two channels but we discuss implications for other channels when we conclude.

We use a static model with two regions that are linked by an international fossil energy market and climate change damages.  Climate policy is implemented only by one region by unilateral carbon pricing.  We derive optimal solutions for emissions reduction and emissions removal in anticipation for this region in anticipation of emissions leakage to the second region.

We find that \emph{removing} a ton of CO$_2$ from the atmosphere unilaterally by one region causes less leakage than \emph{reducing} CO$_2$ emissions by a ton. This is because, in contrast to emissions mitigation, carbon removal does not cause supply-side leakage. Therefore, the optimal CDR subsidy tends to be higher than the optimal tax but lower than marginal climate damages.
Additionally, a country may exploit its market power on the global resource market to change fossil energy price in its favor, depending on the net trade balance. This motive leads to another  strategic wedge between the optimal carbon tax and the optimal carbon removal subsidy.
In certain cases, the difference between optimal carbon tax and CDR subsidy can be expressed as a simple function of the supply side leakage rate.
This paper is, to the best of our knowledge, the first to explore the leakage implications of carbon removal policies in a stylized analytical model (but cf.\ \citealt{Quirion2011} who explore a similar argument for the use of carbon capture and sequestration with fossil fuel combustion in a numerical model).

\section{Model Setup}\label{sec:model}

We consider a two-region economy consisting of one large region $A$ and a rest-of-the-world region $W$, which represents a large number of small countries such that region $W$ acts as a price taker. The regions are populated by a representative household and perfectly competitive firms, which produce a consumption good using labor and fossil energy as inputs. The large region $A$ takes its own contribution to climate change damages into account when setting a domestic carbon tax and a CDR subsidy to maximize welfare. Because countries in $W$ are small, they have no incentives to contribute to climate change mitigation by implementing domestic carbon prices even though they benefit from reduced global damages \citep[see][for a similar small country assumption]{Hoel1992}. 

Representative households maximize utility $u(C^i)$ for $i\in\{A,W\}$, which increases with consumption of a private good $C^i$ with decreasing marginal utility. Consumption goods are produced in each region with the same technology $F(E^i)\Omega(E)$. Here, $E^i$ denotes fossil energy use in region $i$ and $\Omega(E)$ captures environmental damages, which depend on carbon emissions in the atmosphere, $E=E^A+E^W$ \citep{Nordhaus1994,Nordhaus2017}. We assume that $\Omega(0)=1$, $\Omega'(E)<0$, $\Omega''(E)<0$, that is, the fraction $(1-\Omega(E))$ is destroyed by (convex) climate damages. We assume a production function $F$ with positive and decreasing returns to scale due to locally fixed factors such as land or labor. Region $A$ can mitigate its emissions by using less fossil energy $E^A$ or by deploying CDR $R$, that is, mitigation technologies are part of $F$. Costs for removal are weakly convex and given by $h(R)$. 

Finally, fossil energy is sold by competitive fossil energy suppliers at a world market price $p$, maximizing their profits $\pi_R = pE-c(E)$. Extraction costs $c(E)$ are convex and the supply of energy $E$ equals total demand:  $E=E^W+E^A$. Region A owns a fraction $\lambda \in [0,1]$ of the fossil energy suppliers. The first order condition yields 
\begin{align}
    p = c'(E).\label{eq:FOC_fossilsuppliers}
\end{align}

Households own the economy and thus consume the economic output of their region net climate damages, fossil energy purchases and removal costs, and any profits from their fraction of fossil energy producers. 

\section{Discussion of leakage mechanisms}\label{sec:leakage}

\noindent We now discuss the mechanisms by which unilateral emission reductions and deployment of CDR in region $A$ cause leakage. Firms in $W$ maximize profits: 
\begin{align}
    \pi^W &= F(E^W)\Omega(E) - p E^W
\end{align}
utilizing energy up to the point where its price $p$ balances with its marginal productivity:
\begin{align}
    p &= F'(E^W)\Omega \label{eq:focR}
\end{align}
Combining \eqref{eq:FOC_fossilsuppliers} with (\ref{eq:focR}) yields
\begin{align}
    c'(E^A+E^W) &= F'(E^W) \Omega(E^A + E^W - R) \label{eq:EW}
\end{align}

\begin{figure}[tb]
    \centering
\includegraphics[width=0.5\textwidth]{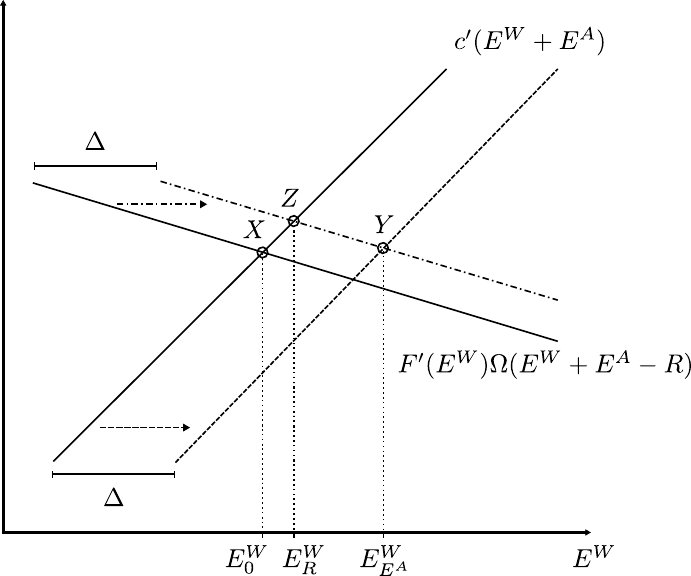}
    \caption{Fossil energy demand $E^W$ in $W$ is such that marginal extraction costs MC are equal to marginal net product $F'(E^W)\Omega$, i.e. marginal benefits MB. This corresponds to Eq.~\eqref{eq:EW}.}
    \label{fig:MC_MB}
\end{figure}

\noindent Figure \ref{fig:MC_MB} explains the energy market equilibrium \eqref{eq:EW} graphically for a simplified case with linear marginal cost and benefit curves from the perspective of region $W$. In the initial equilibrium (point $X$) energy demand is given by $E^W=E_0^W$. If $A$ reduces its demand for fossil energy $E^A$ by some $\Delta$, this has two effects: First, marginal extraction costs fall, shifting the marginal cost curve $c'$ to the right. Second, climate damages fall, shifting the marginal benefit curve $F'(E^W)\Omega$ to the right. Now, marginal costs equal marginal benefits in point $Y$ and
energy demand in $W$ increases from $E_0^W$ to $E_{E^A}^W$. Reducing emissions in $A$ causes supply-side leakage due to the falling price for fossil energy, which stimulates demand in $W$ and free-rider leakage by reducing climate damages, which increases marginal benefits of fossil energy in $W$.

If instead of reducing demand by $\Delta$, region $A$ removes $\Delta$ units of CO$_2$ from the atmosphere, the marginal benefit curve shifts to the right, but marginal extraction costs remain unchanged. Then, the resulting equilibrium is at point $Z$ and emissions in $W$ are only $E_R^W$. Hence, deploying CDR in $A$ causes free-rider leakage.

\noindent We can express the leakage rate of a unilateral emission reduction in region $A$ more precisely by considering that equation \eqref{eq:EW} implicitly determines region $W$'s response function $\phi$ to region $A$'s fuel demand, that is, $E^W(E^A,R)=: \phi(E^A,R)$.

\begin{prop}[Emission reduction leakage rate]  \label{prop:leak}
A unilateral reduction in region $A$'s emissions, $E^A$, leads to an increase in region $W$'s emissions $E^W$ by
\begin{align}
    \frac{d E^W}{d E^A} = \frac{\partial \phi}{\partial E^A} &=  - \left( 1 + \frac{F''(E^W) \Omega}{F'(E^W) \Omega' - c''} \right)^{-1}  \label{eq:dphiEA}
\end{align}
with $-\frac{d E^W}{d E^A}$ denoting the emission reduction leakage rate and $0 < -\frac{d E^W}{d E^A} = -\frac{\partial \phi}{\partial E^A} < 1$
\end{prop}
\begin{proof}
Substituting $E^W=\phi(E^A,R)$  into (\ref{eq:EW}) and taking the total derivative with respect to $E^A$, we obtain $ c''\left(1+\frac{\partial \phi}{\partial E^A}\right) = F''(E^W) \Omega \frac{\partial \phi}{\partial E^A} + F'(E^W) \Omega'\left(1+\frac{\partial \phi}{\partial E^A}\right) $. Re-arranging gives the first result. The second result on the inequality equation follows from $F''(E^W) \Omega<0$, $F'(E^W) \Omega' - c''<0$, implying that  $-1 < \frac{\partial \phi}{\partial E^A} < 0$. 
\end{proof}
The emission reduction leakage rate $-\frac{\partial \phi}{\partial E^A}$ measures how much of the mitigated ton of carbon in region $A$ is off-set by increased energy demand in $W$. Leakage rates are always between 0 and 100\% since $\frac{\partial \phi}{\partial E^A}>-1$. The rate depends on the slopes of the marginal extraction costs and climate damages (cf.~Fig.~\ref{fig:MC_MB}). If, ceteris paribus, $c''$ or $\Omega'$ is large (small) in absolute terms, leakage rates are large (small), too.

Leakage as characterized in Prop.~\ref{prop:leak} thus occurs via the conventional supply-side channel and the free-rider channel. The latter channel is also relevant for CDR:
\begin{prop}[CDR leakage rate] \label{prop:leakCDR} A marginal carbon removal in $A$ affects energy demand, and thus, emissions in $W$ as follows:
\begin{align}
    \frac{d E^W}{d R} = \frac{\partial \phi}{\partial R} &=   \left( 1 + \frac{F''(E^W) \Omega - c''}{F'(E^W) \Omega'} \right)^{-1} > 0  \label{eq:dphiR}
\end{align}
\end{prop}

\begin{proof}
Substituting $E^W=\phi(E^A,R)$ back into (\ref{eq:EW}) and taking the total derivative with respect to $R$, we obtain $ c''\frac{\partial \phi}{\partial R}  = F''(E^W) \Omega\frac{\partial \phi}{\partial R} + F'(E^W) \Omega' (\frac{\partial \phi}{\partial R}-1)  $. The inequality follows from $\frac{F''(E^W) \Omega - c''}{F'(E^W) \Omega'}>0$.
\end{proof}

Carbon removal leakage is induced by reduced climate damages, which increase productivity in $W$ and, thus, demand for (fossil) energy: When damages are flat and $\Omega'$ is small,  $\frac{\partial \phi}{\partial R}$ converges to zero and CDR in region $A$ has almost no effect on fuel use in $W$. When damages are steep and $\Omega'$ is very large, $\frac{\partial \phi}{\partial R}$ converges to one, implying an almost perfect crowding out of CDR by increased emissions abroad. In this case, CDR leads to substantially lower climate damages implying a large increase in fuel demand. We assume convex damages, thus, whether damages are flat or steep depends, amongst others, on whether the model is intended to represent the short run with a relatively low greenhouse gas concentration in the atmosphere (flat damages) or the long run with a relatively high greenhouse gas concentration in the atmosphere (steep damages).

Accordingly, in Fig.~\ref{fig:MC_MB}, a given marginal increase of CDR leads to a small (large) distance between $E_0^W$ and $E_R^W$ for a flat (steep) marginal benefit curve $F'(E^W)\Omega$.
Thus, while CDR induces demand-side leakage through reduced climate damages it does not trigger supply-side leakage. 

Combining \eqref{eq:dphiR} and \eqref{eq:dphiEA} reveals the link between both leakage rates.
\begin{corollary} \label{cor:leak2} For all $(E^A,R)$, emission reduction leakage and CDR leakage are linked by
\begin{align}
    \frac{\partial \phi}{\partial R} &=   \alpha \left( -  \frac{\partial \phi}{\partial E^A}  \right)  \label{eq:dphiR2a}
\end{align}
where $\alpha:=\left( 1 - \frac{c''}{F'(E^W) \Omega'} \right)^{-1}$. The CDR leakage rate is smaller than the emission reduction leakage rate, 
\begin{align}
    \frac{\partial \phi}{\partial R} <  -  \frac{\partial \phi}{\partial E^A}  \label{eq:dphiR2b}
\end{align}
\end{corollary}
\begin{proof}
Equation \eqref{eq:dphiR2a} follows directly from \eqref{eq:dphiEA} and \eqref{eq:dphiR}. The inequality in \eqref{eq:dphiR2b} ctly from (\ref{eq:dphiR2a}) as $0<\left( 1 - \frac{c''}{F'(E^W) \Omega'} \right)^{-1} < 1$.
\end{proof}

\section{Optimal unilateral carbon prices}

The differences in carbon leakage according to Prop.\ 1 and 2 have consequences for region $A$'s optimal carbon tax and removal subsidy. We first derive a 'command-and-control' equilibrium where $A$ sets quantities for fossil energy use and CDR directly, then we use this to solve for the two carbon prices.

\subsection{Command-and-control optimum in \emph{A}}\label{sec:policies_CnC}

The government of $A$ maximizes consumption $C^A$, which we take as a numeraire,%
\footnote{This is equivalent to maximizing utility since the model is static, and utility is monotone in consumption as single determinant.}
i.e.\ it maximizes
\begin{align}
    C^A &= F(E^A)\Omega(E^A + E^W - R) + \lambda \pi_R - p E^A - h(R) \label{eq:objA}
\end{align}
subject to:
\begin{align}
    \pi_R &= p(E^A+E^W) - c(E^A+E^W) \label{eq:pi_R} \\ 
    E^W &= \phi(E^A,R) \label{eq:phi} \\
    p &= c'(E^W + E^A) \label{eq:ToT} 
\end{align}
We substitute \eqref{eq:pi_R} - \eqref{eq:ToT}  into \eqref{eq:objA} and obtain
\begin{align}
    C^A &= F(E^A)\Omega(E^A + \phi(E^A,R) - R) \notag \\  &\hspace{1cm}+ \lambda \left[ c'(E^A + \phi(E^A,R)) (E^A+\phi(E^A,R)) - c(E^A+\phi(E^A,R)) \right] \notag \\
    &\hspace{4cm}-c'(E^A + \phi(E^A,R))E^A-h(R) \label{eq:objA2}
\end{align}
Hence, the government considers not only climate damages that are related to its own choice of fossil energy and removal, it also considers the effect on the share of fossil resource rents, $\lambda$, that are owned by the country.  
With plausible assumptions about functions $F$, $c$ and $h$ at zero and in the limit, any solutions to the maximization of \eqref{eq:objA2} will be interior. %
Maximizing over $(E^A,R)$ gives the first order conditions:\footnote{We assume that functional forms are well-behaved so that the first-order conditions yield an interior maximum. The second order derivatives, however, depend in a non-trivial way on third-order derivatives. The expressions in the Hessian, hence, are too complicated to determine whether it is actually negative definite.}
\begin{align}
  F'(E^A) \Omega - p &= 
   \left(1+\frac{\partial \phi}{\partial E^A}\right)  \left[ -F\Omega'+\Theta
  \right] 
  \label{eq:EA_tot} \\
  h' &= \left( 1+ \alpha \frac{\partial \phi}{\partial E^A} \right)
  \left[ 
      - F\Omega'+  \Theta
      \right] 
      - \Theta
      \label{eq:CDRA_tot}
\end{align}
where we used (\ref{eq:dphiR2a}) for deriving equation \eqref{eq:CDRA_tot} and define $\Theta:=-c''E(\lambda - E^A/E)$, which captures the effect of the resource trade balance on energy and removal choices: when $A$ is a net exporting country, i.e.\ $A$ owns more resource than it utilizes ($\lambda>E^A/E$), $\Theta$ is negative and the right hand side of \eqref{eq:EA_tot} is reduced below the Pigouvian level because an exporter will benefit from extended energy usage.

Leakage due to emission reduction $(\partial \phi/\partial E^A)$ and carbon removal $\alpha (\partial \phi/\partial E^A)$ effectively reduces the impact of marginal climate damages on the optimal choice of domestic emissions and CDR in equations \eqref{eq:EA_tot} and \eqref{eq:CDRA_tot}, respectively.

\subsection{Policy instruments}\label{sec:policies_prices}

We now derive the optimal unilateral carbon tax $\tau$ and CDR subsidy $\varsigma$ of region $A$, and thus, consider a decentralized economy. Firms in $A$ maximize
\begin{align}
    \pi^A &= F(E^A)\Omega(E) - (p+\tau) E^A + \varsigma R - h(R) - w^A L^A
\end{align}
implying the first order conditions:
\begin{align}
    F'(E^A) \Omega &= p + \tau  \label{eq:focAE} \\
    h' &= \varsigma  \label{eq:focAR}
\end{align}
Comparing these first order conditions with the optimality conditions \eqref{eq:EA_tot} and \eqref{eq:CDRA_tot} allows to derive optimal carbon prices for emissions and their removal:
\begin{prop} \label{prop:cpriceleak}
The optimal carbon tax for carbon emissions $\tau^*$ and the optimal subsidy for carbon removal $\varsigma^*$ that maximize region $A$'s welfare are in general not equal. They are given by
\begin{align}
\tau^* &= \left(1+\frac{\partial \phi}{\partial E^A}\right)
  \left[ 
  - F\Omega'
  + \Theta
  \right] \label{eq:taustar}\\
\varsigma^* &= \left( 1+ \alpha \frac{\partial \phi}{\partial E^A} \right)
  \left[ 
      - F\Omega' 
      +\Theta
      \right] 
      - \Theta \label{eq:sigmastar}
\end{align}
with 
\begin{align}
  \Theta &= -c''E(\lambda - E^A/E) \qquad \text{(Resource Trade Balance Effect)}\label{eq:resource_effect}
\end{align}
\end{prop}
\noindent Hence, the optimal carbon tax equals marginal damages $-F\Omega'$ plus a resource trade balance  component $\Theta$.
A net exporting country (i.e.\ $\Theta < 0$) 
has an incentive to lower the carbon tax to benefit from higher resource prices. 
The opposite holds for a country that is a net resource importer. In any case, both marginal damages and resource trade balance component, are adjusted for the emission reduction leakage rate $ 0 <  1+\frac{\partial \phi}{\partial E^A} < 1$. In case of a net zero trade balance, $\lambda = E^A/E$ and $\Theta=0$. 

The optimal CDR subsidy has a similar structure as the optimal carbon tax, but is adjusted for the CDR leakage rate $\left( 1+ \alpha \frac{\partial \phi}{\partial E^A} \right)=\left( 1- \frac{\partial \phi}{\partial R} \right)$ that takes into account increased fossil energy use abroad due to lowered climate damages. Additionally, the resource trade balance effect, $\Theta$, enters in opposite sign again and independent from the leakage effect. The optimal subsidy differs from the optimal carbon tax:
\begin{align}
    \tau^*-\varsigma^* &= \underbrace{(1-\alpha)\frac{\partial \phi}{\partial E^A}(-F\Omega')}_{<0} + \underbrace{\left[
    (1-\alpha)\frac{\partial \phi}{\partial E^A} +1
    \right]}_{>0} \Theta
\end{align}
Without the trade balance effect, $\Theta$, the optimal carbon tax would be lower than the optimal carbon removal subsidy because the latter has a lower carbon leakage rate (expressed by $\alpha<1$). Whether the trade balance effect $\Theta$ is able to reverse this depends on the following cases: 
\begin{enumerate}[a)]
    \item If region A is a net exporter, then $\lambda \geq E^A/E$, $\Theta<0$ and, thus, $ \tau^* < \varsigma^*$. The CDR subsidy is then always larger than the tax on carbon emissions.
    \item If region A is an importer, then $\lambda < E^A/E$, and  $\Theta>0$ is ambiguous. In particular, there is a threshold for $\lambda$ below which $\Theta$ becomes positive. 
    \item If region A has a net zero trade balance, $\lambda = E^A/E$ and $\Theta=0$ and the leakage component is the only relevant determinant, implying that the removal subsidy exceeds the carbon tax rate. 
\end{enumerate}
\noindent In case a), region A benefits from lower carbon taxes due to increasing producer surplus of resource firms; it therefore sets the carbon tax below the removal subsidy to further increase international resource prices. In case b), since region A
imports fossil energy, it has an incentive to use the carbon tax to appropriate the surplus of resource exporters. The carbon tax is therefore higher than in a). When marginal climate damages  $-F\Omega'$ are sufficiently small, the carbon tax might even be higher than the removal subsidy, as the resource trade balance effect dominates the climate damage effect.

The optimal prices simplify significantly, if we assume that region A is self-sufficient in resources (i.e. has a zero trade balance, implying $\Theta=0$) or if A's government does not try to appropriate the resource rent (e.g.~because it wants be a fair player on global resource markets).

\paragraph{Special case.} When the net resource trade balance is zero or when region A takes the global resource price as given, that is, it ignores \eqref{eq:ToT}, the optimal carbon tax and CDR subsidy are
\begin{align}
\hat{\tau} &= -\left(1+\frac{\partial \phi}{\partial E^A}\right)F\Omega' \label{eq:tau_simple}\\
\hat{\varsigma} &=-\left( 1+ \alpha \frac{\partial \phi}{\partial E^A} \right) F\Omega' = -\left( 1- \frac{\partial \phi}{\partial R} \right) F\Omega'\label{eq:sigma_simple}
\end{align}
The result follows directly for the case of a net zero trade balance ($\Theta=0$) from Proposition~\ref{prop:cpriceleak}.\footnote{For the second condition, \eqref{eq:objA} still holds but resource surplus $\pi_R$ and the resource price $p$ is considered exogenous and not influenced by $R$ and $E^A$. Hence, optimizing consumption levels implies optimizing 
$
    C^A = F(E^A)\Omega(E^A + E^W - R) + \zeta - h(R) 
$
where $\zeta:=\lambda \pi_R - p E^A$ is treated as a constant and $\partial\zeta/\partial R=\partial\zeta/\partial E^A= 0$. The rest of the proof is the same as in Proposition \ref{prop:cpriceleak}.
}

With \eqref{eq:tau_simple} and \eqref{eq:sigma_simple} we can put the wedge between removal subsidy and carbon tax into perspective with respect to prior studies on supply-side leakage. Consider the supply-side leakage rate $LR_s:= - \frac{dE^W}{dE^A}_{\vert \Omega'=0}$, which disregards the impacts of climate change.
This is common in this literature \citep[e.g.~in][]{BrangerQuirion2014}, and various numerical or empirical models on supply-side leakage provide estimates of $LR_s$ (which corresponds to our emission reduction leakage rate).
\begin{prop}
If the motive to capture the resource rent is disregarded or region A has a net zero resource trade balance  ($\Theta = 0$), the wedge between the optimal CDR subsidy and the optimal carbon tax depends only on the supply-side leakage rate $LR_s$.
\begin{align}
    \frac{\hat{\varsigma}}{\hat{\tau}} &= \frac{1}{1-LR_s}  \label{eq:leakwedge}
\end{align}
\end{prop}
\begin{proof} Following from \eqref{eq:tau_simple} and \eqref{eq:sigma_simple}, we have to calculate $\frac{1-\frac{\partial \phi}{\partial R}}{1+\frac{\partial \phi}{\partial E^A}}$. With $LR_s:= - \frac{dE^W}{dE^A}_{\vert \Omega'=0}=\frac{c''}{c''-F''(E^W) \Omega}$, we obtain by re-arranging $c''=\frac{F''(E^W)\Omega}{1-LR_s^{-1}}$. Substituting this into (\ref{eq:dphiEA}) and (\ref{eq:dphiR}) we get the result.
\end{proof}
Eq.~\eqref{eq:leakwedge} provides a clear intuition on the optimal wedge between carbon taxes and CDR subsidies, which is determined only by the supply-side leakage rate. If supply side leakage is very high, the optimal CDR subsidy becomes a multiple of the carbon tax, without any upper bound. If supply-side leakage is very small, the CDR subsidy rate converges to the carbon tax rate. %

\citet{sinn_public_2008} argues that supply-side leakage is 100\%. The infinite ratio in \eqref{eq:leakwedge} then suggests not to use the tax ($\hat{\tau}=0$). Proper formal analysis of supply-side leakage emphasized that leakage rates could also be substantially smaller, depending on the size of the climate coalition and demand elasticities, among others \citep{eichner2011carbon}. More realistic estimates of supply-side leakage rate that consider increasing extraction costs of different types of fossil resource find lower rates (in the order of 50\%, putting the optimal subsidy at twice the level of the tax) and, when technological change reduces backstop costs, find also negative leakage rates \citep{fischer2013limits}, putting the optimal subsidy below the emissions tax. \citet{Quirion2011} estimate a supply-side leakage of 37\% (i.e.\ a ratio of 1.6).

\section{Conclusions}\label{sec:conclusion}

Our results challenge the intuition of equal carbon prices for positive and negative emissions by considering the more realistic setting of an internationally fragmented climate policy regime. To the best of our knowledge, our study is the first to shed light on the question of how a pricing policy for CDR in more realistic second-best settings should be designed. 

Our stylized static model generated the following insights: The optimal carbon tax differs from an optimal CDR subsidy because of different carbon leakage and resource balance of trade  motives. With respect to the carbon leakage channel, the optimal removal subsidy tends to be larger than the carbon tax because of lower supply-side leakage on fossil resource markets. This is reinforced for resource owning countries which aim to set removal subsidies higher than carbon taxes to increase resource prices and, thus, surplus of resource producers. This latter effect is, however, ambiguous and depends on the resource trade balance: Net resource exporters aim to increase international resource prices by lower carbon taxes and larger removal subsidies. A resource-poor country may even find it optimal to have a larger carbon tax than a removal subsidy when marginal environmental damages are small -- as the gains from suppressing resource prices may outweigh environmental benefits.

The model captures two channels of carbon leakage: supply-side leakage that works via international energy (or resource) markets and free-rider leakage via climate change damages.  Future research could investigate whether other leakage channels introduce similar asymmetries between emissions tax and removal subsidy.  In particular, of the three commonly investigated channels this paper left specialization leakage unaddressed. Moreover, technology spillovers and input factor markets have been identified as additional leakage channels (as in \citealt{GerlaghKuik2014} and \citealt{BaylisFullertonKarney2014}, respectively) and merit further research.
Future research may also explore further aspects that imply a separate  price for removing carbon versus reducing carbon emissions. Examples include distortive tax systems; geological  storage  sites that are open-access and thus suffer from inefficient dynamic allocation; or when carbon removal is not permanent but small amounts of CO$_2$ leak out of storage sites over time.
Finally, our static setting abstract from dynamic aspects.  But intertemporal leakage (acceleration of fossil resource extraction as in \citealt{sinn_public_2008}) and Hotelling rents, for example, may affect the order of magnitude of the identified effects, and could be explored in a dynamic extension of this work.


\begin{thebibliography}{21}
\newcommand{\enquote}[1]{``#1''}
\providecommand{\natexlab}[1]{#1}
\providecommand{\url}[1]{\texttt{#1}}
\providecommand{\urlprefix}{URL }
\expandafter\ifx\csname urlstyle\endcsname\relax
  \providecommand{\doi}[1]{doi:\discretionary{}{}{}#1}\else
  \providecommand{\doi}{doi:\discretionary{}{}{}\begingroup
  \urlstyle{rm}\Url}\fi
\providecommand{\selectlanguage}[1]{\relax}

\bibitem[{Baylis et~al.(2014)Baylis, Fullerton, and
  Karney}]{BaylisFullertonKarney2014}
Baylis, Kathy, Don Fullerton, and Daniel~H Karney (2014). \enquote{Negative
  leakage.} \emph{Journal of the Association of Environmental and Resource
  Economists}, 1(1/2), 51--73.

\bibitem[{Bohm(1993)}]{Bohm1993}
Bohm, Peter (1993). \enquote{Incomplete international cooperation to reduce CO2
  emissions: alternative policies.} \emph{Journal of Environmental Economics
  and Management}, 24(3), 258--271.

\bibitem[{Branger and Quirion(2014)}]{BrangerQuirion2014}
Branger, Fr{\'e}d{\'e}ric and Philippe Quirion (2014). \enquote{Climate policy
  and the ‘carbon haven’ effect.} \emph{Wiley Interdisciplinary Reviews:
  Climate Change}, 5(1), 53--71.

\bibitem[{Eichner and Pethig(2011)}]{eichner2011carbon}
Eichner, Thomas and R{\"u}diger Pethig (2011). \enquote{Carbon leakage, the
  green paradox, and perfect future markets.} \emph{International Economic
  Review}, 52(3), 767--805.

\bibitem[{Fischer and Salant(2013)}]{fischer2013limits}
Fischer, Carolyn and Stephen~W Salant (2013). \enquote{Limits to limiting
  greenhouse gases: Intertemporal leakage, spatial leakage, and negative
  leakage.} \emph{Working Paper}.

\bibitem[{Fuss et~al.(2018)Fuss, Lamb, Callaghan, Hilaire, Creutzig, Amann,
  Beringer, Garcia, Hartmann, Khanna, Luderer, Nemet, Rogelj, Smith, Vicente,
  Wilcox, Dominguez, and Minx}]{fuss_negative_2018}
Fuss, Sabine, William~F. Lamb, Max~W. Callaghan, Jérôme Hilaire, Felix
  Creutzig, Thorben Amann, Tim Beringer, Wagner de~Oliveira Garcia, Jens
  Hartmann, Tarun Khanna, Gunnar Luderer, Gregory~F. Nemet, Joeri Rogelj, Pete
  Smith, José Luis~Vicente Vicente, Jennifer Wilcox, Maria del Mar~Zamora
  Dominguez, and Jan~C. Minx (2018). \enquote{Negative emissions—{Part} 2:
  {Costs}, potentials and side effects.} \emph{Environmental Research Letters},
  13(6), 063002. Publisher: IOP Publishing.

\bibitem[{Gerlagh and Kuik(2014)}]{GerlaghKuik2014}
Gerlagh, Reyer and Onno Kuik (2014). \enquote{{Spill or leak? Carbon leakage
  with international technology spillovers: A CGE analysis}.} \emph{Energy
  Economics}, 45, 381--388.

\bibitem[{Groom and Venmans(2021)}]{GroomVenmans2021}
Groom, Ben and Frank Venmans (2021). \enquote{The social value of offsets.}
  Presented at the {Virtual Seminar on Climate Economics}, {Federal Reserve
  Bank of San Francisco}.

\bibitem[{Hoel(1991)}]{Hoel1991}
Hoel, Michael (1991). \enquote{Global environmental problems: the effects of
  unilateral actions taken by one country.} \emph{Journal of environmental
  economics and management}, 20(1), 55--70.

\bibitem[{Hoel(1992)}]{Hoel1992}
Hoel, Michael (1992). \enquote{{International environment conventions: the case
  of uniform reductions of emissions}.} \emph{Environmental and Resource
  Economics}, 2(2), 141--159.

\bibitem[{IPCC(2018)}]{ipcc_global_2018}
IPCC (2018). \enquote{Global {Warming} of 1.5 {Degree} {Celsius}: {IPCC}
  {Special} {Report} on the {Impacts} of {Global} {Warming} of 1.5 {Degree}
  {Celsius}.} \emph{Intergovernmental Panel on Climate Change, Geneva,
  Switzerland, Report.}

\bibitem[{Jakob et~al.(2014)Jakob, Steckel, and
  Edenhofer}]{JakobSteckelEdenhofer2014}
Jakob, Michael, Jan~Christoph Steckel, and Ottmar Edenhofer (2014).
  \enquote{{Consumption-versus production-based emission policies}.}
  \emph{Annu. Rev. Resour. Econ.}, 6(1), 297--318.

\bibitem[{Kalkuhl et~al.(2022)Kalkuhl, Franks, Gruner, Lessmann, and
  Edenhofer}]{KalkuhlFranksGrunerEtAl2022}
Kalkuhl, Matthias, Max Franks, Friedemann Gruner, Kai Lessmann, and Ottmar
  Edenhofer (2022). \enquote{Carbon pricing when carbon removal is not
  permanent.} Presented at the {EAERE 2022 Annual Conference}, Rimini.

\bibitem[{Lemoine(2020)}]{Lemoine2020}
Lemoine, Derek (2020). \enquote{Incentivizing Negative Emissions Through Carbon
  Shares.} Tech. rep., National Bureau of Economic Research.

\bibitem[{Minx et~al.(2018)Minx, Lamb, Callaghan, Fuss, Hilaire, Creutzig,
  Amann, Beringer, Garcia, Hartmann, Khanna, Lenzi, Luderer, Nemet, Rogelj,
  Smith, Vicente, Wilcox, and Dominguez}]{minx_negative_2018}
Minx, Jan~C., William~F. Lamb, Max~W. Callaghan, Sabine Fuss, Jerome Hilaire,
  Felix Creutzig, Thorben Amann, Tim Beringer, Wagner de~Oliveira Garcia, Jens
  Hartmann, Tarun Khanna, Dominic Lenzi, Gunnar Luderer, Gregory~F. Nemet,
  Joeri Rogelj, Pete Smith, Jose Luis~Vicente Vicente, Jennifer Wilcox, and
  Maria del Mar~Zamora Dominguez (2018).
  {\selectlanguage{english}\enquote{Negative emissions-{Part} 1: {Research}
  landscape and synthesis.}} \emph{Environmental Research Letters}, 13(6).
  Publisher: IOP Publishing Ltd.

\bibitem[{Nemet et~al.(2018)Nemet, Callaghan, Creutzig, Fuss, Hartmann,
  Hilaire, Lamb, Minx, Rogers, and Smith}]{nemet_negative_2018}
Nemet, Gregory~F., Max~W. Callaghan, Felix Creutzig, Sabine Fuss, Jens
  Hartmann, Jerome Hilaire, William~F. Lamb, Jan~C. Minx, Sophia Rogers, and
  Pete Smith (2018). {\selectlanguage{english}\enquote{Negative
  emissions-{Part} 3: {Innovation} and upscaling.}} \emph{Environmental
  Research Letters}, 13(6). Publisher: IOP Publishing Ltd.

\bibitem[{Nordhaus(1994)}]{Nordhaus1994}
Nordhaus, William~D (1994). \emph{Managing the global commons: the economics of
  climate change}, vol.~31. MIT press Cambridge, MA.

\bibitem[{Nordhaus(2017)}]{Nordhaus2017}
Nordhaus, William~D (2017). \enquote{Revisiting the social cost of carbon.}
  \emph{Proceedings of the National Academy of Sciences}, 114(7), 1518--1523.

\bibitem[{Quirion et~al.(2011)Quirion, Rozenberg, Sassi, and
  Vogt-Schilb}]{Quirion2011}
Quirion, Philippe, Julie Rozenberg, Olivier Sassi, and Adrien Vogt-Schilb
  (2011). \enquote{How {$CO_2$} Capture and Storage can Mitigate Carbon
  Leakage.} {FEEM Working Paper} No. 15.2011, available at SSRN:
  http://dx.doi.org/10.2139/ssrn.1763165.

\bibitem[{Siebert(1979)}]{siebert_environmental_1979}
Siebert, Horst (1979). \enquote{Environmental {Policy} in the
  {Two}-{Country}-{Case}.} \emph{Zeitschrift für Nationalökonomie / Journal
  of Economics}, 39(3/4), 259--274. Publisher: Springer.

\bibitem[{Sinn(2008)}]{sinn_public_2008}
Sinn, Hans-Werner (2008). \enquote{Public policies against global warming: a
  supply side approach.} \emph{International Tax and Public Finance}, 15(4),
  360--394.

\end{thebibliography}

\end{document}